\documentclass[5p,twocolumn]{elsarticle}
\usepackage{graphicx,latexsym}
\usepackage{dcolumn}
\usepackage{amssymb,amsmath,bm}
\usepackage{subfigure}
\usepackage{braket}
\usepackage{siunitx}
\usepackage{array}
\usepackage{float}
\usepackage[nopar]{lipsum}
\usepackage{tikz}
\usepackage{caption}
\usepackage{multirow}
\usepackage{bigstrut}
\usepackage[super]{nth}

\newcommand{\angstrom}{\text{\normalfont\AA}}
\usepackage{hyperref}
\hypersetup{
	pdfnewwindow=true,       
	colorlinks=true,         
	linkcolor=black,          
	citecolor=red,          
	filecolor=magenta,       
	urlcolor=black           
}

\usepackage[normalem]{ulem}

\def\sec#1{Sec.\ \ref{#1}}
\def\eq#1{Eq.\ (\ref{#1})}
\def\fig#1{Fig.\ \ref{#1}}
\def\tab#1{Tab.\ \ref{#1}}

\journal{}

\begin{document}

\begin{frontmatter}


\title{Study of BC$_{14}$N-bilayer graphene: Effects of atomic spacing and interatomic interaction between B and N atoms} 


\author[a1,a2]{Nzar Rauf Abdullah}
\ead{nzar.r.abdullah@gmail.com}
\address[a1]{Division of Computational Nanoscience, Physics Department, College of Science, 
             University of Sulaimani, Sulaimani 46001, Kurdistan Region, Iraq}
\address[a2]{Computer Engineering Department, College of Engineering, Komar University of Science and Technology, Sulaimani 46001, Kurdistan Region, Iraq}

\author[a1]{Hunar Omar Rashid}

\author[a4]{Vidar Gudmundsson}
\ead{vidar@hi.is}
\address[a4]{Science Institute, University of Iceland, Dunhaga 3, IS-107 Reykjavik, Iceland}


\begin{abstract}
	
We study the effects of an attractive interaction between the boron (B) and the nitrogen (N) atoms doped in a bilayer graphene (BLG), BC$_{14}$N, on the electronic, the thermal and the optical properties for two different types of a doping process: First, both the B and the N atoms are doped in the same layer while the other layer is undoped. Second, the B and N atoms are doped in both layers. An attractive interaction between the B and N atoms does not influence the interlayer interaction in the first case, while it does in the second case. We find that the strong B-N attractive interaction in one layer induces metallic behavior due to the crossing of the valence band and the Fermi energy, while the strong attractive interaction between both layers induces a semiconductor property arising from the emergence a bandgap. 
We therefore confirm that the metallic-like BLG is not a good material for thermal devices because it has a low figure of merit, while we notice that the semiconductor-like BLG has a high Seebeck coefficient and figure of merit as well as a low thermal conductivity. 
The strong attractive interaction of the B-N atoms between the layers gives rise to a prominent peak to appear in dielectric function, the excitation and the absorption spectra in the low energy, visible range, while a very weak peak is seen in the case of a strong attractive interaction between the B and N doped in one layer. Controlling the B and N atomic configurations in the BLG may help to improve the material for use in both thermoelectric and optoelectronic devices.

\end{abstract}


\end{frontmatter}

\section{Introduction} 

The structure of materials is described by their chemical compositions and specific arrangements of atoms in a crystal lattice. The electronic structure can be changed by controlling: First, intrinsic factors, such as the interatomic distance or the bonding in the atomic arrangement \cite{McCann_2013, Thompson2020}. Second, extrinsic factors, such as an electric field, the temperature, the pressure, and impurities \cite{Oostinga2008, Gonnelli2015, Yankowitz2016}. 
The two-dimensional BLG,
constituted by two stacked graphene layers, is a material in which it's physical properties can be controlled by both intrinsic and extrinsic factors. BLG has attracted much interest because it has given indications for an exceptionally high charge mobility, like is seen in single monolayer graphene \cite{Zhang2009}. 
The high value of mobility is a result of a decreased electron–phonon interaction that leads to a significantly lower carrier scattering. This indicates that BLG a very attractive material for high-speed transistors \cite{Liao2010}. In addition, an inverse relationship between the carrier mobility and the bandgap in graphene has been confirmed~\cite{doi:10.1063/1.4792142, doi:10.1021/acsomega.0c01676}, and it 
has been shown that bandgap tuning by extrinsic factors changes the carrier mobility. The changes in carrier mobility affect the entire physical properties \cite{PhysRevB.80.235402, Song2012, Radamson2017}.

Recently, the band structure of BLG has been investigated with respect to the extrinsic factors, such as the presence of substitutional B and/or N doping, using density functional theory (DFT) with van der Waals correction \cite{Alattas2018}.
It has been shown that the introduction of B-N pairs into bilayer graphene can be used to create a considerable band gap if the B atom is doped in one layer and the N atom is doped into the other layer, but B-N pairs doped into one of the layers of BLG hardly modifies the band dispersion.
Furthermore, the attractive interaction between the B and the N atoms doped in BLG has been investigated and it has been shown that the attractive interaction can induce more flexible mechanical properties of the system and lead to a decreased optical response \cite{ABDULLAH2020100740}. A conversion of the stacking orientation of bilayer graphene due to the interaction of the BN-dopants has been reported \cite{abdullah2021conversion}.
There was demonstrated that in the presence of a repulsive interaction between the B and N atoms, the AA-stacking is converted to a AB-stacking with a more stable structure. It enhances mechanical properties, such as leading to a higher Young modulus, the ultimate strength and stress, fracture
strength comparing to the AA-stacked BN-codoped BLG \cite{abdullah2021role}.
In the current work, we consider a BN-codoped BLG forming BC$_{14}$N, where the BN pair is doped either in one layer or in both layers. We confirm the results of previous studies that reported a small bandgap 
if the BN pair doped in one layer, but a larger bandgap when the B atom is doped in one layer and the N atom is doped into the other layer. This is caused by the attractive interaction between the layers. In addition we report the thermal and optical properties for both cases here. 

In \sec{Sec:Model} the BLG structure is briefly over-viewed. In \sec{Sec:Results} the main 
results achieved are analyzed. In \sec{Sec:Conclusion} the conclusion of the modeling is presented.

\section{Computational Tools}\label{Sec:Model}

The Quantum espresso (QE) package for solving the Kohn–Sham DFT equations is used to study the electronic and optical properties of the system where the plane-wave pseudopotential method has been implemented \cite{Giannozzi_2009, giannozzi2017advanced, abdullah2020properties}. Thermal properties of the systems are calculated by the Boltztrap package based on Boltzman transport equations \cite{Madsen2006}.
The vdW interaction in our model is taken into account and we assume the DFT-D technique with vdw-DF exchange–correlation functionals \cite{Berland_2015}. We can thus take into account the long-range electron correlations. 
The accuracy of these pseudoptentials for studying structural and electronic characteristics have been verified in our previous works \cite{ABDULLAH2021114644, ABDULLAH2020114556, ABDULLAH2020126807}.  
The cutoffs of plane-wave energy is chosen to be $816$~eV with the $20 \times 20 \times 2$ of the $k$-grid sampling in the first Brillouin zone for the fully relaxed structures. The structures are considered to have reached stability with $10^{-6}$~eV/$\angstrom$ as the tolerance of maximal force on each atoms.
In the SCF calculations, the same number of $k$-points in the grid have been used, while in the 
DOS calculations the number of $k$-points in the grid is increased to  $70 \times 70 \times 2$.

\section{Results and discussion}\label{Sec:Results}
 
\begin{figure}[htb]
	\centering
	\includegraphics[width=0.35\textwidth]{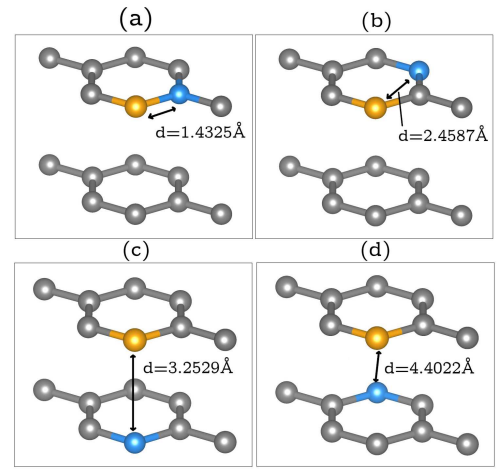}
	\caption{Bilayer graphene with B and N atoms, BC$_{14}$N, doped in the top layer with the B-N distance of $d = 1.43$ (a), and $d = 2.45$ $\angstrom$ (b), and the B atom is doped in the top layer while the N atom is doped in the bottom layer with the B-N distance of $3.25$ (c) and $4.40 \, \angstrom$ (d), respectively. The C, B and N atoms are colored in gray, yellow, and blue.}
	\label{fig01}
\end{figure}

The systems under investigation consists of a 2 × 2 supercell BLG with B
and N dopant atoms, BC$_{14}$N, where the B and N atoms are either doped in one or both layers.
We consider four atomic configurations based on the distance between the B and N atoms, 
and all four configurations are presented in \fig{fig01} where the C, B and N atoms are colored in gray, yellow, and blue. To visualize the structures, we use the XCrySDen software \cite{KOKALJ1999176}.
In the first two atomic configurations (a and b), both the B and N atoms are doped in only one layer (top layer), while in the last two atomic configurations (c and d), the B and N atoms are doped in both layers \cite{rani2013designing}. 
In the first configuration, the distance between the B and the N atoms is equal to the length of a BN-bond, which is $d = 1.43$ $\angstrom$ (see \fig{fig01}(a)), while in the second structure the distance between the B and the N atom is increased to $2.45 \, \angstrom$ (b). The B and N atoms in these two structures are doped in the top layer, while the bottom layer is undoped. In order to increase the distance between the B and N atoms, we now consider the B atom is doped in the top layer while the N atom is doped in the bottom layer with the distance $3.25 \, \angstrom$ (c) and $4.40 \, \angstrom$ (d). We should mention that the B and N atoms in both layers (the $3^{rd}$ and the $4^{th}$ type of configurations) are doped in the same atomic site of hexagonal structure of graphene (A- or B-site) which leads to preservation of the AA-stacking structure of BLG in our systems here. If we would assume that the boron atom is doped at an A-site of the top layer and the nitrogen atom in a B-site of the bottom layer, the AA-stacking of BLG is changed to a AB-stacking due to the repulsive interaction between the B and N atoms as was recently reported \cite{abdullah2021conversion}.

The inter layer distance, $h$, the distance between the B and N atoms, $d$, the interaction energy, $\Delta E$, and the bandgap, $\Delta_g$, are presented in \tab{table_one}.
\begin{table}[h]
	\centering
	\begin{center}
		\caption{\label{table_one} Interlayer distance, $h$, distance between B and N atoms, $d$, interaction energy, $\Delta E$, and bandgap, $\Delta_g$, for the four BN-codoped BLG structures.}
		\begin{tabular}{|l|l|l|l|l|l|l|l|}\hline
			Structure	 & h ($\angstrom $)   & d ($\angstrom $)   & $\Delta E$ (eV)    & $ \Delta_g$  (eV) \\ \hline
			a	&  3.57   &   1.43    &     -5.32    & 	 0.15 	\\
			b   &  3.57   &   2.45    &     -4.93    &   0.14	\\
			c	&  3.21   &   3.25    &     -4.04    &   0.21   \\
			d	&  3.35   &   4.4     &     -3.18    &   --     \\   \hline 
	\end{tabular}	\end{center}
\end{table}
The interaction between the B and N dopant atoms in BLG structure can be calculated via 
\begin{equation}
	\Delta E = E_2 - E_0 + 2 \times E_1,
\end{equation}
where E$_0$, E$_1$ and E$_2$ are the total energies of the systems
with zero, one, and two dopant atoms, respectively. 
The negative sign of interaction energy shown in 
\tab{table_one} indicates that there is an attractive interaction between the B and N atoms in all four considered structures here \cite{ABDULLAH2020126350}. It seems that the attractive interaction is inversely proportional with the distance between the B and N atoms. The attractive interaction is the strongest and weakest for the structure with $d = 1.43$ (a), and $4.4 \, \angstrom$ (d), respectively. 
It has been shown that the interaction strength is almost zero when the separation distance between the the B and N atoms is greater than or equal to 4.0 $\angstrom$ in the absence of the vdw-DF when the GGA or the LDA are assumed \cite{doi:10.1063/1.4742063, ABDULLAH2020103282}. But in the presence of the vdw-DF of our system, the 
attractive interaction is not zero for the largest distance between the B and N atoms, $d = 4.4 \, \angstrom$ (d) which is due to presence of the van der Waals interaction taken into account 
via the vdw-DF. 

In addition, the calculated value for the interlayer spacing, $h$, of the first (a) and the second (b) structure, where both the B and N atoms are doped into the top layer, confirms that the attractive interactions between the B and N atom do not influence the interlayer spacing of these two structures, $h = 3.57 \, \angstrom$, as their interlayer spacing is almost equal to the interlayer spacing of pure BLG, $h = 3.58 \, \angstrom$ \cite{Chung2002, Razado-Colambo2018}. 
On the other hand, the interlayer spacing of the last two structure (c), and (d), is smaller than that of pure BLG indicating, that the attractive interaction between the B and N atoms leads to a reduction of the interlayer spacing compared to pure BLG.

\begin{figure}[htb]
	\centering
	\includegraphics[width=0.45\textwidth]{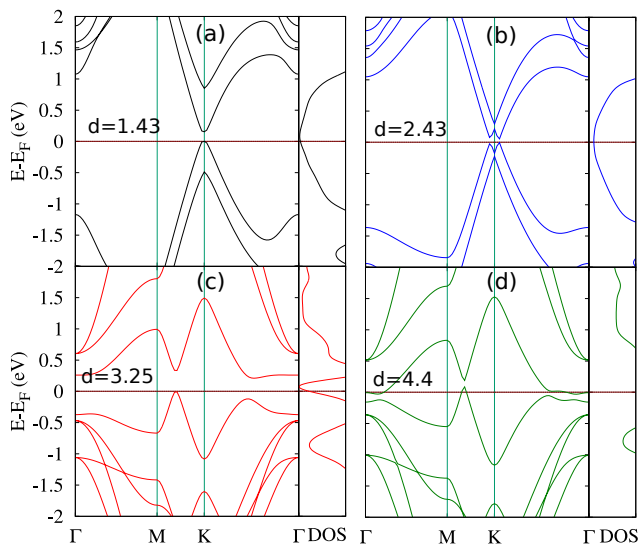}
	\caption{Electronic energy dispersion of the BN-codoped BLG where the distance between the B and N dopant atoms is $d=1.43$ (a), $2.45$ (b), $3.25$ (c), and $4.4\, \angstrom$ (d) with their total DOS at right side of the dispersion energy. The Fermi energy is set to zero.}
	\label{fig02}
\end{figure}

The energy dispersion of all structures with their DOS are shown in \fig{fig02}.
It has been shown that pure AA-stacked BLG has multiple linear dispersion bands. 
The linear dispersion bands are mainly due to the electronic interlayer coupling that is
suppressed by the Pauli repulsion between the graphene layers \cite{C2NR30823A}.
The first observation of our results is that linear dispersion around the K-point is not seen anymore in the vicinity of the Fermi energy. In the first (a) and second (b) structures multiple dispersion bands are still seen due to the presence of 
both B and N dopant atoms in the top layer, where the interaction between dopant atoms does not affect the interlayer interaction, but the interaction between the B and N atoms in the third (c) and fourth (d) structures removes several dispersion bands and forms instead 
two ``mexican hats'' around the K-point. This is similar to the energy dispersion of BLG subject to a perpendicular electric field \cite{Chuang2013, PhysRevLett.99.216802}.
Furthermore, the largest bandgap is found for the structure that contains the strongest interlayer interaction arsing from the attractive interaction between the B and N atoms in different layers (see \fig{fig02}(c)).

The bandgap values of all structures are presented in \tab{table_one}. The structures with B-N distance, $d = 1.43$ and $4.4 \, \angstrom$ have a metallic property as the top of the valence band slightly crosses the Fermi energy, while the structures with  $d = 2.45$ and $3.25 \, \angstrom$ have a semiconductor behavior with a direct and an indirect bandgap, respectively.

\begin{figure}[htb]
 	\centering
 	\includegraphics[width=0.45\textwidth]{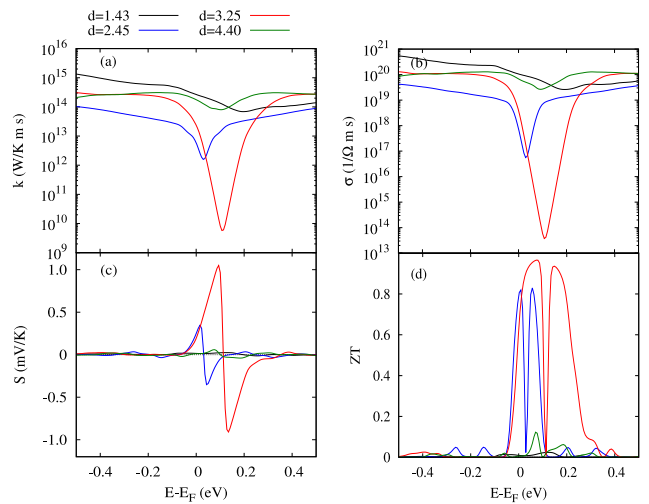}
 	\caption{Electronic thermal conductivity, $k$, (a), electrical conductivity, $\sigma$, (b), Seebeck coefficient, $S$, (c), and figure of merit (d) versus energy are plotted for all four structures.}
 	\label{fig03}
\end{figure}

The thermoelectric properties of the systems are demonstrated in \fig{fig03} at the temperature 
$T = 100$~K where the phonon contribution to the transport properties is almost inactivated \cite{PhysRevB.87.241411}. To investigate the electronic thermal properties, we use the BoltzTraP software for semi-classical transport coefficients. The code
uses a mesh of band energies and is interfaced to the QE package.
 
A device with high figure of merit should have high Seebeck coefficient and electrical conductivity with low thermal conductivity \cite{C7EE02007D, ABDULLAH2020126578}. 
The lack of thermal electrons in the bandgap range leads to a decrease in the electronic thermal conductivity \cite{RASHID2019102625, abdullah2019thermoelectric}. As s result, the electronic thermal conductivity is lower around the Fermi energy in the case of a larger bandgap. We can thus see that the electronic thermal conductivity is well suppressed, but the Seebeck coefficient is enhanced around the Fermi energy for the structure with the largest bandgap, $0.21$~eV (see \fig{fig03}a and c, red color), where a strong interlayer interaction is present in the system. Consequently, the highest value of figure of merit, $ZT$, is found for the structure with the $0.21$~eV bandgap, where the interlayer distance is $3.25  \, \angstrom$.

BLG is characterized by good transparency and optical response.
The optical properties of BLG are different for both the directions of the applied electric field, parallel, E$_{\parallel}$ or perpendicular, E$_{\perp}$ to the surface of the structures, due to the hexagonal symmetry in the system.
The optical properties of the BLG can be computed by the QE package, where the QE code generates data for the frequency dependent real part, $\varepsilon_1(\omega)$, and the imaginary part, $\varepsilon_2(\omega)$, of the dielectric function, $\varepsilon(\omega) = \varepsilon_1(\omega) + i \, \varepsilon_2(\omega)$. 
One can thus use these data sets to compute the related optical characteristics. 
For instance, the refractive index and the excitation spectra can be computed via \cite{dressel2002electrodynamics, RANI201428}
\begin{equation}
	n(\omega) = \Big(  \frac{\sqrt{\varepsilon_1^2 + \varepsilon_2^2} + \varepsilon_1}{2} \Big)^{1/2}
\end{equation}
and 
\begin{equation}
	k(\omega) = \Big(  \frac{\sqrt{\varepsilon_1^2 + \varepsilon_2^2} - \varepsilon_1}{2} \Big)^{1/2} .
\end{equation}
The reflectivity at normal incidence of a EM wave on the systems can be computed from 
$n$ and $k$ by 
\begin{equation}
	R(\omega) = \frac{(n-1)^2 + k^2}{(n+2)^2 + k^2}.
\end{equation} 
In addition, the absorption spectra can be calculated using 
\begin{equation}
	\alpha(\omega) = \frac{2 k \omega}{c\hbar},
	\label{alpha}
\end{equation}
where $c$ is the speed of light in vacuum, and $\omega$ is in energy units. 

Were, we show how the separation between the B and N dopant atoms influences the optical properties, such as the dielectric function, the absorption spectra, the reflectivity, 
and the electron energy loss function. 

\begin{figure}[htb]
	\centering
	\includegraphics[width=0.5\textwidth]{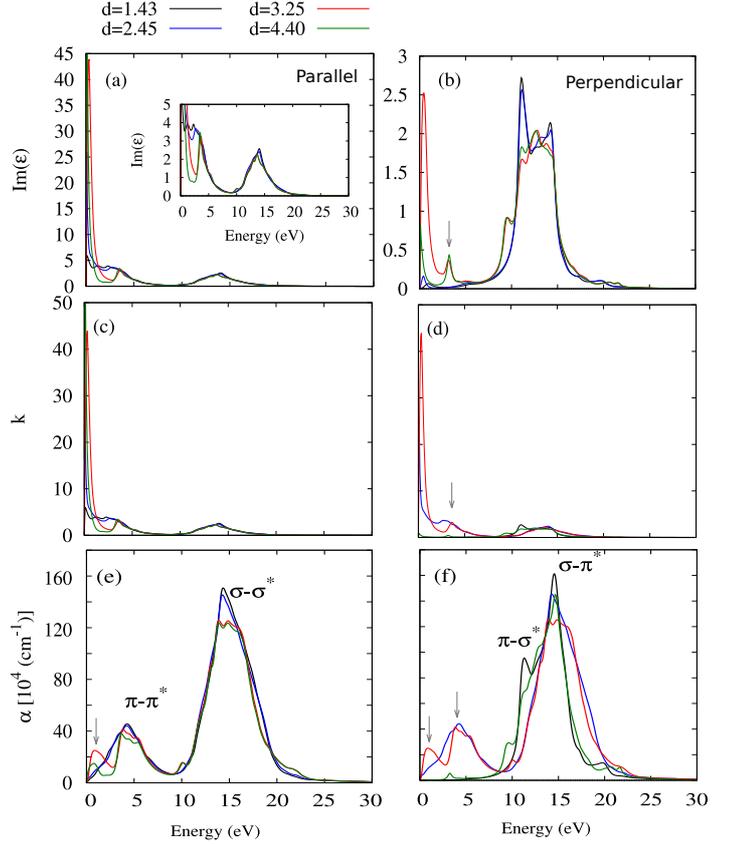}
	\caption{The imaginary part of the dielectric function, Im($\epsilon$), the excitation spectra, $k$, and the absorption coefficient, $\alpha$, are shown for both parallel (a,c,e) and perpendicular (b,d,f) electric fields, respectively.}
	\label{fig04}
\end{figure}

In \fig{fig04}, the imaginary part of dielectric function, Im($\varepsilon$) or $\varepsilon_2$, excitation spectra, $k$, and absorption coefficient, $\alpha$, are shown for both the parallel (a,c,e) and the perpendicular (b,d,f) electric field (E-field), respectively. 
It has been shown that two main peaks for pure BLG in the imaginary part of the dielectric function at $3.95$ and $13.87$~eV are formed in the case of E$_{\parallel}$ by the $\pi$ to $\pi^*$ and the $\sigma$ to $\sigma^*$ transitions, respectively.
In addition, two main peaks are generated by the transitions from the $\sigma$ to $\pi^*$ at $11.22$~eV and the $\pi$ to $\sigma^*$ at $14.26$~eV in the case of E$_{\perp}$ \cite{NATH2015691, ABDULLAH2020100740}. The expected anisotropy between the two different polarizations is clearly seen. 

In the presence of the BN dopant atoms with different distance between the B and N atoms both peaks are also formed with a slight shift to lower energy for both direction of the E-field. The inset of \fig{fig04}(a) is nothing but the Im($\varepsilon$) for E$_{\parallel}$ for a smaller scale of the y-axis. The Im($\varepsilon$) spectrum for both polarizations of the E-field is dominated by a very intense peak structure at low frequencies in the visible energy range, energy $< 1.0$~eV, indicating a transition due to the opening up of a bandgap.
An extra peak at $3.12$~eV (gray arrow) appears due to the formation of ``mexican hats'' in the dispersion energy around the K-point for the E$_{\perp}$.
The same scenario is true for the excitation spectra directly related to the absorption spectra via \eq{alpha}.  
In addition, a pronounced peak in the absorption spectra at low energy for the structure with the largest bandgap is formed for both polarizations of the E-field (gray arrow). The peak is again due to the opening of the bandgap caused by the strong interlayer interaction between the B and N atoms  (see \fig{fig04}(e and f)).

\begin{figure}[htb]
	\centering
	\includegraphics[width=0.5\textwidth]{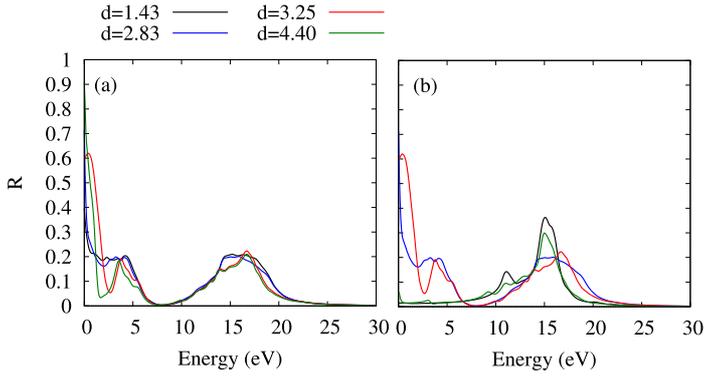}
	\caption{Reflectivity function for E$_{\parallel}$ (a), and E$_{\perp}$ (b).}
	\label{fig05}
\end{figure}

After showing the absorption spectra, we now present the reflectivity in \fig{fig05} for both 
E$_{\parallel}$ (a) and E$_{\perp}$ (b). The reflectivity is computed
from the frequency dependent real and imaginary parts of the refractive index.
It has been shown that two main peaks in the reflectivity spectra appear for pure graphene 
at $4.5$ and $15.0$~eV for E$_{\parallel}$, and one prominent peak at $15.3$~eV for E$_{\perp}$ \cite{NATH2014275}. 
In N-doped BLG, the N atom increases the free electron density in the systems, that’s why a significant increase of the reflectivity is expected while the B-doping decreases free electron density in the system leading to a notable decrease in the reflectivity of the system \cite{doi:10.1021/nn506074u}. In presence of both the B and the N dopant atoms of our systems, the generated free electron density depends on the interaction between the B and the N atoms. It seems that a dipole-dipole interaction between the B and the N atoms is generated leading to a balance of the free electron density in the system \cite{abdullah2021role}. The charge is transferred between the B and the N atoms instead of supplying/receiving charge to/from the BLG. Consequently, a big change in intensity of the two main peaks of reflectivity is not see for E$_{\parallel}$ irrespective of the distance between the B and the N atoms or the interaction between the B and the N atoms. 
The dipole-dipole interaction between the B and the N atoms can be influenced by the E$_{\perp}$ leading to a variation of the density of free electrons in the system and changes in the peak intensity at $15.0$~eV.        
In both polarization cases, for E$_{\parallel}$ and E$_{\perp}$,  a strong peak is formed at low energy for the structure with strong interlayer distance, $d = 3.25\, \angstrom$ (red color).

The electron energy-loss spectra (EELS) is determined by inelastic scattering of fast electrons in the systems. The energy distribution of all the inelastically scattered electrons gives
information about the local environment of the atomic electrons, which in turn relates to the physical and chemical properties of the systems. It seems that the two main peaks in EELS are not very sensitive on the distance between the B and N atoms in the case of E$_{\parallel}$, while the most intense peak at $15-20$~eV is shifted to the higher energy for E$_{\perp}$. Most importantly, an extra peak below $3.0$~eV for E$_{\parallel}$ and two peaks below $9.0$~eV for E$_{\perp}$ are caused by a plasma resonance of the valence electrons, which seems to be strong for the system with interlayer interaction due to the B and N atoms.
\begin{figure}[htb]
	\centering
	\includegraphics[width=0.5\textwidth]{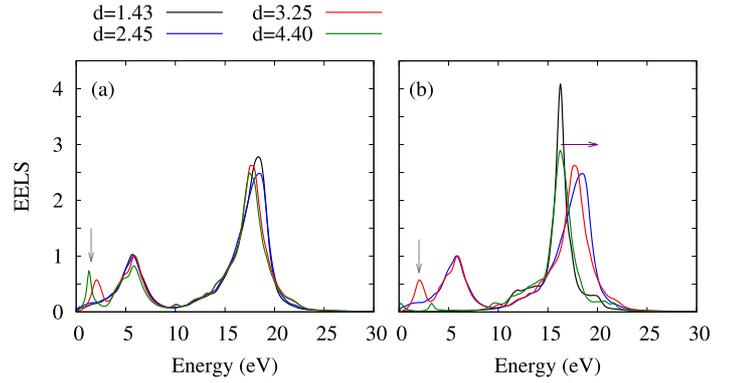}
	\caption{Electron energy lose spectra, EELS, for all four considered structures shown in \fig{fig01} for the case of E$_{\parallel}$ (a), and E$_{\perp}$ (b).}
	\label{fig06}
\end{figure}

\section{Conclusions}\label{Sec:Conclusion}

Overall, electronic, thermal and optical properties of a bilayer graphene (BLG) with attractive interaction between dopant B and N atoms within a layer, or between both layers have been investigated. In this work, a density functional modeling implemented through the Quantum Espresso package has been used. We demonstrate that a weak attractive interaction between the B and N dopant atoms induces a semiconductor-like BLG structure when B and N atoms are doped in the same layer and the other layer is undoped. In contrast, a strong attractive interaction between the layers due to the B and N atoms causes a semiconductor-like BLG. In both cases, a relatively larger band gap was found leading to a good thermoelectric materials. A strong peak in the excitation and the absorption spectra was observed indicating a good optical material in the visible region of 
the electromagnetic spectrum.

\section{Acknowledgment}
This work was financially supported by the University of Sulaimani and 
the Komar University of Science and Technology. 
The computations were performed on resources provided by the Division of Computational 
Nanoscience at the University of Sulaimani. 
 


\begin{thebibliography}{50}
	\expandafter\ifx\csname url\endcsname\relax
	\def\url#1{\texttt{#1}}\fi
	\expandafter\ifx\csname urlprefix\endcsname\relax\def\urlprefix{URL }\fi
	\expandafter\ifx\csname href\endcsname\relax
	\def\href#1#2{#2} \def\path#1{#1}\fi
	
	\bibitem{McCann_2013}
	E.~McCann, M.~Koshino,
	\href{https://doi.org/10.1088}{} Reports on Progress in Physics 76~(5) (2013)
	056503.
	\newblock \href {https://doi.org/10.1088/0034-4885/76/5/056503}
	{\path{doi:10.1088/0034-4885/76/5/056503}}.
	\newline\urlprefix\url{https://doi.org/10.1088}
		
		\bibitem{Thompson2020}
		J.~J.~P. Thompson, D.~Pei, H.~Peng, H.~Wang, N.~Channa, H.~L. Peng, A.~Barinov,
		N.~B.~M. Schr{\"o}ter, Y.~Chen, M.~Mucha-Kruczy{\'{n}}ski,
		\href{https://doi.org/10.1038/s41467-020-17412-0}{} Nature Communications 11~(1) (2020) 3582.
		\newblock \href {https://doi.org/10.1038/s41467-020-17412-0}
		{\path{doi:10.1038/s41467-020-17412-0}}.
		\newline\urlprefix\url{https://doi.org/10.1038/s41467-020-17412-0}
		
		\bibitem{Oostinga2008}
		J.~B. Oostinga, H.~B. Heersche, X.~Liu, A.~F. Morpurgo, L.~M.~K. Vandersypen,
		\href{https://doi.org/10.1038/nmat2082}{} Nature Materials 7~(2) (2008) 151--157.
		\newblock \href {https://doi.org/10.1038/nmat2082}
		{\path{doi:10.1038/nmat2082}}.
		\newline\urlprefix\url{https://doi.org/10.1038/nmat2082}
		
		\bibitem{Gonnelli2015}
		R.~S. Gonnelli, F.~Paolucci, E.~Piatti, K.~Sharda, A.~Sola, M.~Tortello, J.~R.
		Nair, C.~Gerbaldi, M.~Bruna, S.~Borini,
		\href{https://doi.org/10.1038/srep09554}{} Scientific Reports 5~(1) (2015) 9554.
		\newblock \href {https://doi.org/10.1038/srep09554}
		{\path{doi:10.1038/srep09554}}.
		\newline\urlprefix\url{https://doi.org/10.1038/srep09554}
		
		\bibitem{Yankowitz2016}
		M.~Yankowitz, K.~Watanabe, T.~Taniguchi, P.~San-Jose, B.~J. LeRoy,
		\href{https://doi.org/10.1038/ncomms13168}{} Nature Communications 7~(1) (2016)
		13168.
		\newblock \href {https://doi.org/10.1038/ncomms13168}
		{\path{doi:10.1038/ncomms13168}}.
		\newline\urlprefix\url{https://doi.org/10.1038/ncomms13168}
		
		\bibitem{Zhang2009}
		Y.~Zhang, T.-T. Tang, C.~Girit, Z.~Hao, M.~C. Martin, A.~Zettl, M.~F. Crommie,
		Y.~R. Shen, F.~Wang, \href{https://doi.org/10.1038/nature08105}{} Nature
		459~(7248) (2009) 820--823.
		\newblock \href {https://doi.org/10.1038/nature08105}
		{\path{doi:10.1038/nature08105}}.
		\newline\urlprefix\url{https://doi.org/10.1038/nature08105}
		
		\bibitem{Liao2010}
		L.~Liao, Y.-C. Lin, M.~Bao, R.~Cheng, J.~Bai, Y.~Liu, Y.~Qu, K.~L. Wang,
		Y.~Huang, X.~Duan, \href{https://doi.org/10.1038/nature09405}{} Nature 467~(7313)
		(2010) 305--308.
		\newblock \href {https://doi.org/10.1038/nature09405}
		{\path{doi:10.1038/nature09405}}.
		\newline\urlprefix\url{https://doi.org/10.1038/nature09405}
		
		\bibitem{doi:10.1063/1.4792142}
		J.~Wang, R.~Zhao, M.~Yang, Z.~Liu, Z.~Liu,
		\href{https://doi.org/10.1063/1.4792142}{} The Journal of Chemical Physics 138~(8)
		(2013) 084701.
		\newblock \href {http://arxiv.org/abs/https://doi.org/10.1063/1.4792142}
		{\path{arXiv:https://doi.org/10.1063/1.4792142}}, \href
		{https://doi.org/10.1063/1.4792142} {\path{doi:10.1063/1.4792142}}.
		\newline\urlprefix\url{https://doi.org/10.1063/1.4792142}
		
		\bibitem{doi:10.1021/acsomega.0c01676}
		S.~H. Mir, V.~K. Yadav, J.~K. Singh,
		\href{https://doi.org/10.1021/acsomega.0c01676}{}
		ACS Omega 5~(24) (2020) 14203--14211.
		\newblock \href {http://arxiv.org/abs/https://doi.org/10.1021/acsomega.0c01676}
		{\path{arXiv:https://doi.org/10.1021/acsomega.0c01676}}, \href
		{https://doi.org/10.1021/acsomega.0c01676}
		{\path{doi:10.1021/acsomega.0c01676}}.
		\newline\urlprefix\url{https://doi.org/10.1021/acsomega.0c01676}
		
		\bibitem{PhysRevB.80.235402}
		W.~Zhu, V.~Perebeinos, M.~Freitag, P.~Avouris,
		\href{https://link.aps.org/doi/10.1103/PhysRevB.80.235402}{} Phys. Rev. B 80 (2009) 235402.
		\newblock \href {https://doi.org/10.1103/PhysRevB.80.235402}
		{\path{doi:10.1103/PhysRevB.80.235402}}.
		\newline\urlprefix\url{https://link.aps.org/doi/10.1103/PhysRevB.80.235402}
		
		\bibitem{Song2012}
		H.~S. Song, S.~L. Li, H.~Miyazaki, S.~Sato, K.~Hayashi, A.~Yamada, N.~Yokoyama,
		K.~Tsukagoshi, \href{https://doi.org/10.1038/srep00337}{} Scientific Reports 2~(1) (2012) 337.
		\newblock \href {https://doi.org/10.1038/srep00337}
		{\path{doi:10.1038/srep00337}}.
		\newline\urlprefix\url{https://doi.org/10.1038/srep00337}
		
		\bibitem{Radamson2017}
		H.~H. Radamson, \href{https://doi.org/10.1007/978-3-319-48933-9_48}{Graphene},
		Springer International Publishing, Cham, 2017, pp. 1--1.
		\newblock \href {https://doi.org/10.1007/978-3-319-48933-9_48}
		{\path{doi:10.1007/978-3-319-48933-9_48}}.
		\newline\urlprefix\url{https://doi.org/10.1007/978-3-319-48933-9_48}
		
		\bibitem{Alattas2018}
		M.~Alattas, U.~Schwingenschl{\"o}gl,
		\href{https://doi.org/10.1038/s41598-018-35671-2}{} Scientific Reports 8~(1) (2018) 17689.
		\newblock \href {https://doi.org/10.1038/s41598-018-35671-2}
		{\path{doi:10.1038/s41598-018-35671-2}}.
		\newline\urlprefix\url{https://doi.org/10.1038/s41598-018-35671-2}
		
		\bibitem{ABDULLAH2020100740}
		N.~R. Abdullah, H.~O. Rashid, A.~Manolescu, V.~Gudmundsson,
		\href{http://www.sciencedirect.com/science/article/pii/S246802302030732X}{} Surfaces and Interfaces 21 (2020) 100740.
		\newblock \href {https://doi.org/https://doi.org/10.1016/j.surfin.2020.100740}
		{\path{doi:https://doi.org/10.1016/j.surfin.2020.100740}}.
		\newline\urlprefix\url{}
		
		\bibitem{abdullah2021conversion}
		N.~R. Abdullah, H.~O. Rashid, C.-S. Tang, A.~Manolescu, V.~Gudmundsson,
		arXiv preprint
		arXiv:2101.00462 (2021).
		
		\bibitem{abdullah2021role}
		N.~R. Abdullah, H.~O. Rashid, C.-S. Tang, A.~Manolescu, V.~Gudmundsson,
		arXiv preprint arXiv:2102.09543
		(2021).
		
		\bibitem{Giannozzi_2009}
		P.~Giannozzi, S.~Baroni, N.~Bonini, M.~Calandra, R.~Car, C.~Cavazzoni,
		D.~Ceresoli, G.~L. Chiarotti, M.~Cococcioni, I.~Dabo, A.~D. Corso,
		S.~de~Gironcoli, S.~Fabris, G.~Fratesi, R.~Gebauer, U.~Gerstmann,
		C.~Gougoussis, A.~Kokalj, M.~Lazzeri, L.~Martin-Samos, N.~Marzari, F.~Mauri,
		R.~Mazzarello, S.~Paolini, A.~Pasquarello, L.~Paulatto, C.~Sbraccia,
		S.~Scandolo, G.~Sclauzero, A.~P. Seitsonen, A.~Smogunov, P.~Umari, R.~M.
		Wentzcovitch,
		\href{https://doi.org/10.1088}{
			}, Journal of Physics: Condensed Matter 21~(39)
		(2009) 395502.
		\newblock \href {https://doi.org/10.1088/0953-8984/21/39/395502}
		{\path{doi:10.1088/0953-8984/21/39/395502}}.
		\newline\urlprefix\url{https://doi.org/10.1088}
			
			\bibitem{giannozzi2017advanced}
			P.~Giannozzi, O.~Andreussi, T.~Brumme, O.~Bunau, M.~B. Nardelli, M.~Calandra,
			R.~Car, C.~Cavazzoni, D.~Ceresoli, M.~Cococcioni, et~al., Journal of
			Physics: Condensed Matter 29~(46) (2017) 465901.
			
			\bibitem{abdullah2020properties}
			N.~R. Abdullah, Properties of bilayer graphene-like si $ \_ $\{$2$\}$ $ c $ \_
			$\{$14$\}$ $ semiconductor using first-principle calculations, arXiv preprint
			arXiv:2012.02699 (2020).
			
			\bibitem{Madsen2006}
			G.~K.~H. Madsen, D.~J. Singh,
			\href{http://www.sciencedirect.com/science/article/pii/S0010465506001305}{} Computer Physics
			Communications 175~(1) (2006) 67--71.

			
			\bibitem{Berland_2015}
			K.~Berland, V.~R. Cooper, K.~Lee, E.~Schröder, T.~Thonhauser, P.~Hyldgaard,
			B.~I. Lundqvist, \href{https://doi.org/10.1088/0034-4885/78/6/066501}{} Reports on Progress in Physics 78~(6) (2015) 066501.
			\newblock \href {https://doi.org/10.1088/0034-4885/78/6/066501}
			{\path{doi:10.1088/0034-4885/78/6/066501}}.

			
			\bibitem{ABDULLAH2021114644}
			N.~R. Abdullah, M.~T. Kareem, H.~O. Rashid, A.~Manolescu, V.~Gudmundsson,
			\href{https://www.sciencedirect.com/science/article/pii/S1386947721000266}{} Physica E: Low-dimensional Systems
			and Nanostructures 129 (2021) 114644.
			\newblock \href {https://doi.org/https://doi.org/10.1016/j.physe.2021.114644}
			{\path{doi:https://doi.org/10.1016/j.physe.2021.114644}}.

			
			\bibitem{ABDULLAH2020114556}
			N.~R. Abdullah, H.~O. Rashid, C.-S. Tang, A.~Manolescu, V.~Gudmundsson,
			\href{http://www.sciencedirect.com/science/article/pii/S1386947720316246}{} Physica E:
			Low-dimensional Systems and Nanostructures (2020) 114556\href
			{https://doi.org/https://doi.org/10.1016/j.physe.2020.114556}
			{\path{doi:https://doi.org/10.1016/j.physe.2020.114556}}.

			
			\bibitem{ABDULLAH2020126807}
			N.~R. Abdullah, H.~O. Rashid, C.-S. Tang, A.~Manolescu, V.~Gudmundsson,
			\href{http://www.sciencedirect.com/science/article/pii/S0375960120306745}{} Physics Letters A 384~(32) (2020)
			126807.
			\newblock \href
			{https://doi.org/https://doi.org/10.1016/j.physleta.2020.126807}
			{\path{doi:https://doi.org/10.1016/j.physleta.2020.126807}}.

			
			\bibitem{KOKALJ1999176}
			A.~Kokalj,
			\href{http://www.sciencedirect.com/science/article/pii/S1093326399000285}{}
			Journal of Molecular Graphics and Modelling 17~(3) (1999) 176--179.
			\newblock \href {https://doi.org/10.1016/S1093-3263(99)00028-5}
			{\path{doi:10.1016/S1093-3263(99)00028-5}}.

			
			\bibitem{rani2013designing}
			P.~Rani, V.~Jindal, Rsc
			Advances 3~(3) (2013) 802--812.
			
			\bibitem{ABDULLAH2020126350}
			N.~R. Abdullah, H.~O. Rashid, M.~T. Kareem, C.-S. Tang, A.~Manolescu,
			V.~Gudmundsson,
			\href{http://www.sciencedirect.com/science/article/pii/S0375960120301602}{} Physics Letters
			A 384~(12) (2020) 126350.
			\newblock \href {https://doi.org/10.1016/j.physleta.2020.126350}
			{\path{doi:10.1016/j.physleta.2020.126350}}.

			
			\bibitem{doi:10.1063/1.4742063}
			N.~Al-Aqtash, K.~M. Al-Tarawneh, T.~Tawalbeh, I.~Vasiliev,
			\href{https://doi.org/10.1063/1.4742063}{} Journal of Applied Physics
			112~(3) (2012) 034304.
			\newblock \href {http://arxiv.org/abs/https://doi.org/10.1063/1.4742063}
			{\path{arXiv:https://doi.org/10.1063/1.4742063}}, \href
			{https://doi.org/10.1063/1.4742063} {\path{doi:10.1063/1.4742063}}.

			
			\bibitem{ABDULLAH2020103282}
			N.~R. Abdullah, D.~A. Abdalla, T.~Y. Ahmed, S.~W. Abdulqadr, H.~O. Rashid,
			\href{http://www.sciencedirect.com/science/article/pii/S2211379720317496}{} Results in Physics 18 (2020) 103282.
			\newblock \href {https://doi.org/https://doi.org/10.1016/j.rinp.2020.103282}
			{\path{doi:https://doi.org/10.1016/j.rinp.2020.103282}}.

			
			\bibitem{Chung2002}
			D.~D.~L. Chung, \href{https://doi.org/10.1023/A:1014915307738}{} Journal of Materials Science 37~(8) (2002) 1475--1489.
			\newblock \href {https://doi.org/10.1023/A:1014915307738}
			{\path{doi:10.1023/A:1014915307738}}.

			
			\bibitem{Razado-Colambo2018}
			I.~Razado-Colambo, J.~Avila, D.~Vignaud, S.~Godey, X.~Wallart, D.~P. Woodruff,
			M.~C. Asensio, \href{https://doi.org/10.1038/s41598-018-28402-0}{} Scientific Reports 8~(1) (2018) 10190.
			\newblock \href {https://doi.org/10.1038/s41598-018-28402-0}
			{\path{doi:10.1038/s41598-018-28402-0}}.

			
			\bibitem{C2NR30823A}
			J.~Hargrove, H.~B.~M. Shashikala, L.~Guerrido, N.~Ravi, X.-Q. Wang,
			\href{http://dx.doi.org/10.1039/C2NR30823A}{} Nanoscale 4 (2012) 4443--4446.
			\newblock \href {https://doi.org/10.1039/C2NR30823A}
			{\path{doi:10.1039/C2NR30823A}}.

			
			\bibitem{Chuang2013}
			Y.-C. Chuang, J.-Y. Wu, M.-F. Lin,
			\href{https://doi.org/10.1038/srep01368}{} Scientific Reports 3~(1)
			(2013) 1368.
			\newblock \href {https://doi.org/10.1038/srep01368}
			{\path{doi:10.1038/srep01368}}.

			
			\bibitem{PhysRevLett.99.216802}
			E.~V. Castro, K.~S. Novoselov, S.~V. Morozov, N.~M.~R. Peres, J.~M. B.~L. dos
			Santos, J.~Nilsson, F.~Guinea, A.~K. Geim, A.~H.~C. Neto,
			\href{https://link.aps.org/doi/10.1103/PhysRevLett.99.216802}{}
			Phys. Rev. Lett. 99 (2007) 216802.
			\newblock \href {https://doi.org/10.1103/PhysRevLett.99.216802}
			{\path{doi:10.1103/PhysRevLett.99.216802}}.

			
			\bibitem{PhysRevB.87.241411}
			S.~{Yi\ifmmode \breve{g}\else {\u g}\fi{}en}, V.~Tayari, J.~O. Island, J.~M.
			Porter, A.~R. Champagne,
			\href{https://link.aps.org/doi/10.1103/PhysRevB.87.241411}{} Phys. Rev. B 87 (2013)
			241411.
			\newblock \href {https://doi.org/10.1103/PhysRevB.87.241411}
			{\path{doi:10.1103/PhysRevB.87.241411}}.

			
			\bibitem{C7EE02007D}
			G.~J. Snyder, A.~H. Snyder, \href{http://dx.doi.org/10.1039/C7EE02007D}{}
			Energy Environ. Sci. 10 (2017) 2280--2283.
			\newblock \href {https://doi.org/10.1039/C7EE02007D}
			{\path{doi:10.1039/C7EE02007D}}.

			
			\bibitem{ABDULLAH2020126578}
			N.~R. Abdullah, G.~A. Mohammed, H.~O. Rashid, V.~Gudmundsson,
			\href{http://www.sciencedirect.com/science/article/pii/S037596012030445X}{} Physics Letters A 384~(24) (2020) 126578.
			\newblock \href
			{https://doi.org/https://doi.org/10.1016/j.physleta.2020.126578}
			{\path{doi:https://doi.org/10.1016/j.physleta.2020.126578}}.

			
			\bibitem{RASHID2019102625}
			H.~O. Rashid, N.~R. Abdullah, V.~Gudmundsson,
			\href{http://www.sciencedirect.com/science/article/pii/S2211379719317140}{}
			Results in Physics 15 (2019) 102625.
			\newblock \href {https://doi.org/10.1016/j.rinp.2019.102625}
			{\path{doi:10.1016/j.rinp.2019.102625}}.

			
			\bibitem{abdullah2019thermoelectric}
			N.~R. Abdullah, C.-S. Tang, A.~Manolescu, V.~Gudmundsson,
			Nanomaterials 9~(5) (2019) 741.
			
			\bibitem{dressel2002electrodynamics}
			M.~Dressel, G.~Gr{\"u}ner, Electrodynamics of solids: optical properties of
			electrons in matter (2002).
			
			\bibitem{RANI201428}
			P.~Rani, G.~S. Dubey, V.~Jindal,
			\href{http://www.sciencedirect.com/science/article/pii/S1386947714001374}{} Physica E:
			Low-dimensional Systems and Nanostructures 62 (2014) 28--35.
			\newblock \href {https://doi.org/10.1016/j.physe.2014.04.010}
			{\path{doi:10.1016/j.physe.2014.04.010}}.

			
			\bibitem{NATH2015691}
			P.~Nath, D.~Sanyal, D.~Jana,
			\href{http://www.sciencedirect.com/science/article/pii/S1567173915000887}{} Current Applied Physics 15~(6) (2015) 691 -- 697.
			\newblock \href {https://doi.org/https://doi.org/10.1016/j.cap.2015.03.011}
			{\path{doi:https://doi.org/10.1016/j.cap.2015.03.011}}.

			
			\bibitem{NATH2014275}
			P.~Nath, S.~Chowdhury, D.~Sanyal, D.~Jana,
			\href{https://www.sciencedirect.com/science/article/pii/S000862231400205X}{} Carbon 73 (2014) 275--282.
			\newblock \href {https://doi.org/https://doi.org/10.1016/j.carbon.2014.02.064}
			{\path{doi:https://doi.org/10.1016/j.carbon.2014.02.064}}.

			
			\bibitem{doi:10.1021/nn506074u}
			Y.~Tison, J.~Lagoute, V.~Repain, C.~Chacon, Y.~Girard, S.~Rousset, F.~Joucken,
			D.~Sharma, L.~Henrard, H.~Amara, A.~Ghedjatti, F.~Ducastelle,
			\href{https://doi.org/10.1021/nn506074u}{} ACS Nano 9~(1) (2015) 670--678, pMID:
			25558891.
			\newblock \href {http://arxiv.org/abs/https://doi.org/10.1021/nn506074u}
			{\path{arXiv:https://doi.org/10.1021/nn506074u}}, \href
			{https://doi.org/10.1021/nn506074u} {\path{doi:10.1021/nn506074u}}.

			
		\end{thebibliography}

\end{document}